\documentclass{elsart}
\usepackage{amssymb}
\usepackage{epsfig}
\newcommand{\lapx}{\mbox{\raisebox{-4pt}{$\,\buildrel<\over\sim\,$}}} 
\newcommand{\gapx}{\mbox{\raisebox{-4pt}{$\,\buildrel>\over\sim\,$}}} 

\begin{document}

\begin{frontmatter}
\title{Intrinsic Coulomb blockade in multi-wall carbon nanotubes}
\author[l1]{R. Egger}
\author[l2]{, A.O. Gogolin}
\address[l1]{Institut f\"ur Theoretische Physik IV, 
Heinrich-Heine-Universit\"at, \\ D-40225 D\"usseldorf, Germany}
\address[l2]{Department of Mathematics, Imperial College, 180 Queen's Gate,\\
London SW7 2BZ, United Kingdom}
\begin{abstract}
Carbon nanotubes provide a new class of molecular wires that
display new and exciting mesoscopic transport properties.
We provide a detailed theoretical description for transport in
multi-wall nanotubes, where both disorder and strong interactions
are important.  The interplay of both aspects leads to a particularly effective
intrinsic Coulomb blockade for tunneling.  The relation to recent
experiments is discussed.
\end{abstract}


\end{frontmatter}

\section{Introduction}
\label{Intro}

Transport in molecular wires has received a lot of
attention during the past decade.  Besides 
fundamental interest, much of the relevance of this field comes
from potential applications in the realm of
molecular electronics \cite{joachim}.
In this article, we will focus on one specific class
of molecular wires, namely {\sl carbon nanotubes} \cite{dekker99}.
Nanotubes provide a remarkable and exciting arena for
mesoscopic transport phenomena involving strong 
electron correlations.  
The primary quantity theoretically analyzed below
is the energy-dependent tunneling density of states (TDOS)
for tunneling into the nanotube,
$\nu(E) \sim \,{\rm Re}\,\int_0^\infty dt \, e^{iEt}
 \langle \psi^{}(t)\psi^\dagger(0)\rangle$,
where energies are measured relative to the Fermi level $E_F$
(we put $\hbar=1$).
 The energy dependence
of the TDOS directly governs the (nonlinear) conductance
of a nanotube connected to an STM tip or to metallic leads 
via bad contacts (tunnel junctions).  It also determines 
the intrinsic conductance in the presence of strong
impurity backscattering or weak links.  
From a more  fundamental point of view, the TDOS
provides precious information about the importance
of Coulomb interactions and electronic correlations 
in such a molecular wire.
Recent experiments on individually contacted single-wall nanotubes (SWNTs),
which are composed of a single wrapped graphite sheet, have 
convincingly established the {\sl ballistic} (essentially defect-free)
 nature of electronic transport in SWNTs over distances of the
order 1 $\mu$m and beyond \cite{nt1,nt2}.
Due to their small radius $R\approx 1$ to 2 nm, SWNTs are
characterized by strong transverse momentum quantization,
and under normal circumstances only two 
spin-degenerate transport bands
are present at the Fermi level (assuming a metallic SWNT).
In such a one-dimensional (1D) conductor, the electron-electron
interactions are expected to be of crucial importance.
In fact, the field-theoretical analysis \cite{egger97,kane97}
predicted the breakdown of conventional Fermi-liquid theory.
On not too low energy scales, in practice meaning that the temperature
should be above the sub-mK regime, a Luttinger liquid
(LL) phase should emerge in SWNTs.  The LL is the generic phase of interacting
1D electrons, and is characterized by the absence
of Landau quasiparticles, implying a smeared Fermi surface.
In addition,  a LL exhibits
spin-charge separation, electron fractionalization, and 
anomalous transport properties \cite{ll2}.
Recent SWNT experiments \cite{ll1,ll1a,ll1b,ll1c,ll1d} reported
clear evidence for the elusive LL behavior 
of 1D interacting fermions.  
These charge transport experiments have measured  the
TDOS for tunneling into the SWNT.  According to LL theory, 
the energy dependence of the TDOS is a power law,
with  an exponent that explicitly depends on whether one tunnels into the end
or into the middle of the SWNT.  The predicted exponents
have by now been observed experimentally with good
precision \cite{ll1a,ll1b,ll1c,ll1d}.
For a review on the status of theory and experiment regarding electronic
correlation effects in SWNTs, see Ref.~\cite{review}. 

The situation is more complex and controversial for
multi-wall nanotubes (MWNTs), 
which are the focus of this article.
Existing experimental observations for MWNTs 
do not seem to easily fit into the framework of well-established theories
for disordered electrons.  The reason appears
to be linked to the presence of strong electron-electron interactions
as will be exemplified below for
the case of the TDOS. 
The structure of this article is as follows.
In Sec.~\ref{sec2}, we summarize basic transport
properties of MWNTs and the experimental situation.
In Sec.~\ref{sec3},  phenomenological Coulomb blockade
theory is reviewed, which is then given a microscopic
justification based on 
a nonlinear $\sigma$ model calculation.
The theoretical predictions
for the TDOS, in particular the numerical solution of the Coulomb blockade
equations is discussed in Sec.~\ref{sec4}.
Finally, some conclusions can be found in Sec.~\ref{sec5}.

\section{Electronic transport in MWNTs}
\label{sec2}

Multi-wall nanotubes 
are composed of several (typically ten) concentrically arranged
graphite shells, with outermost radius of the order $R\approx 5$ to 10~nm
and length $L$ up to several 100~$\mu$m.  It is rather obvious that 
the main difference between MWNTs and SWNTs, apart from
the larger radii of MWNTs, should come from the presence of inner shells.
Because of the large radius, 
the transverse quantization energy is only $v_F/R\approx 0.2$~eV,
where the Fermi velocity is $v_F=8\times 10^5$ m/sec,
and therefore one needs to be careful about the position of the
Fermi level.  While in an undoped MWNT, electron-hole
symmetry enforces $E_F=0$, in basically all tubes studied so
far a rather strong doping effect was present, 
$|E_F|\approx 0.3$ to $0.5$~eV \cite{sc4}.  
The physical origin of the doping is largely open,
but may be a result of charge transfer from oxygen or from the substrate
or the attached leads.  This means in practice that
typically 20 spin-degenerate subbands are present 
(instead of only two as in SWNTs), 
and therefore doped MWNTs correspond to {\sl multi-channel
molecular quantum wires}.  Looking for the moment only at the outermost shell,
the bandstructure of a perfect (clean) tube corresponds to a 
Dirac ``light cone'', $E(\vec{k})=v_F |\vec{k}|$, around each of
the two gapless  K points \cite{dekker99}, 
with $\vec{k}=(k,k_\perp)$ and quantized transverse momentum, $k_\perp=n/R$,
where $n=-N,\ldots,N$ and $N=[k_F R]$. 
The number $M=2N+1 \approx 10$ of spin-degenerate  1D subbands at each
K point arising from periodic
boundary conditions around the circumference is 
determined by the doping (Fermi) level via $k_F=|E_F|/v_F$. 
The $n$th subband is then characterized by a separate Fermi velocity,
$v_n=v_F \sqrt{1-(n/k_F R)^2}$,
 and Fermi momentum,  $k_n = k_F v_n/v_F$.
For clarity, we focus on doped MWNTs throughout this article, 
where $M\gg 1$.

With one exception \cite{frank}, available experiments agree that 
electronic transport in MWNTs is not ballistic (as in SWNTs)  but 
{\sl diffusive} \cite{langer,sc1,sc2,jap,liu,graug,kim,sc3,finn,sc5,Lu}.
Estimates for the mean free paths differ substantially in 
different studies, and seem to depend on many aspects, e.g.~MWNT fabrication, 
purification and preparation, as well as the energy regime probed
experimentally.
To mention some of the experimental evidence for
diffusive behaviors, there are typical
weak localization features, universal conductance fluctuations
 and $h/2e$ oscillations in the magnetoconductance.
These experiments also show that electronic transport
proceeds only through the outermost shell (which is contacted
by external leads), unless that shell has been intentionally
damaged.  There are a number of theoretical arguments
supporting this observation \cite{kane,egger99,prl}, and we shall
assume an effective single-shell model in what follows.
Inner shells then cause a screening of the electron-electron
interaction potential. 
For a computation of the TDOS, the latter effect, as well as the
spin and K point degeneracy, 
can be absorbed by a suitable renormalization of the 
electron-electron interaction potential $U_0(\vec{q})$ \cite{egger99}.
Therefore we may simplify the computation and 
consider only spinless electrons at
one K point.  In addition,
since different shells always have incommensurate lattices
due to different curvature or helicity,
a quasiperiodic ionic potential from inner shells acts on
outermost-shell electrons.  The effect of such a potential
is expected to be similar to a random potential described by a 
mean free path $\ell=v_F\tau$.
Moreover, true disorder imposed by imperfections, substrate
inhomogeneities, tube processing or defects is very likely present.
Together with the inner-shell potential, this may explain the
reported mean free paths,  $\ell \leq 100$~nm.
Estimating the localization length as $\xi \approx 4 M \ell$
using the standard Thouless argument, 
electronic transport at low energy scales, $E\tau \ll 1$,
is therefore diffusive (but not localized)
 for not exceedingly long doped MWNTs.  However, for
$E\tau \gg 1$, there is a ballistic regime, where the
Luttinger liquid picture \cite{egger99} is appropriate
again, with minor modifications. 
 In this article, we will mainly focus on the more complex
energy regime $E\tau \ll 1$, i.e.~assume a sufficiently
dirty MWNT.  In fact, we impose the condition $\ell \lapx 2\pi R$ such that
transport around the circumference can be considered as
diffusive. The opposite limit $\ell \gapx 2\pi R$ has recently been addressed 
in Ref.~\cite{gl1}.  

At low energy scales, $E<v_F/R$, many groups have by now
experimentally observed pronounced {\sl zero-bias anomalies} in the TDOS
of an individual MWNT \cite{sc2,liu,graug,kim,sc3,finn,Lu}.  
Most of these experimental results are described by a power-law 
TDOS, $\nu(E)\propto E^\alpha$,
 just as in a LL, with exponents clustering around $\alpha\approx 
0.3 \pm 0.1$ \cite{sc2,graug,sc3,finn}. 
Remarkably, this value is of the same order of magnitude as the
exponents in  a SWNT and hence the interpretation in terms of a LL may seem
obvious. One exception to this 
result has been reported in Ref.~\cite{liu},
where $\alpha\approx 0.04\pm 0.02$.
Occasionally, also logarithmic dependencies have been observed \cite{Lu}
in MWNT bundles, where probably the electron-electron interaction
is externally screened.  Such logarithmic dependencies could
be explained by tunneling into an effectively 2D diffusive system
with weak Coulomb interactions \cite{altshuler}.
Furthermore, the TDOS at the end of the MWNT, while still
of power-law form, is characterized by a doubling of the
boundary exponent, $\alpha_{\rm end}=2 \alpha$.
Unfortunately, it appears to be difficult to explain these findings 
by Luttinger liquid theory, at least for the majority
of the quoted experimental studies.  Although
LL theory can be extended
to ballistic multi-mode wires with inner-shell screening
 \cite{egger99}, the presence
of many subbands in a doped MWNT inevitably implies rather
small exponents.  Even optimistic estimates yield exponents that
are at least one order of magnitude smaller than observed.
For that reason, below we address the role of disorder for
the zero-bias anomaly.

Building upon our original paper \cite{prl}, we provide a theoretical
description for the TDOS of MWNTs within the general
framework of {\sl Coulomb blockade theory} \cite{sct}. 
We focus on sufficiently low energy scales $E < v_F/R$, where
it is sufficient to take a fixed number $M$ of subbands,
and thereby ignore
van Hove singularities associated with the opening 
of new 1D subbands as energy is varied.
As is shown below, on intermediate energy scales, an apparent
power law suppression of the TDOS is found, which is distinct
from the LL power laws of a ballistic system.
This {\sl intrinsic Coulomb blockade} phenomenon
arises because of the suppression of
tunneling into an {\sl strongly interacting disordered metal}.
Because of strong interactions, a perturbative Altshuler-Aronov-Lee
(AAL) approach \cite{altshuler} is not possible. 
The corresponding nonperturbative problem for 2D systems 
 has been studied in Refs.~\cite{finkel,belitz,levitov,kopietz,ka}. 
Very recently, besides our own paper \cite{prl},
the 1D case has attracted considerable interest in the
theory community \cite{gl1,roll,kopietz2}.

\section{Intrinsic Coulomb Blockade}
\label{sec3}

The key ingredient in Coulomb blockade theory is the probability $P(E)$ that 
a tunneling electron excites electromagnetic modes
 with energy $E$ in the system \cite{sct}.  The theory
is meaningful if these modes are harmonic, and then $P(E)$
directly determines the TDOS according to a relation
first explicitly given in Ref.~\cite{roll},
\begin{equation} \label{tdossct}
\frac{\nu(E)}{\nu_0} = \int_{-\infty}^\infty dE' \frac{1+\exp(-E/k_B T)}{1+
\exp(-E'/k_B T)} P(E-E') \;,
\end{equation} 
where $\nu_0$ is the non-interacting DOS.
The probability $P(E)$ is expressed as
\begin{equation}\label{pe}
P(E) = \frac{1}{\pi} {\rm Re} \int_0^\infty dt \exp[iEt+J(t)] \;,
\end{equation}
with the phase correlation function
\begin{equation} \label{jf}
J(t)  =  \int_0^\infty d\omega \frac{I(\omega)}{\omega}
\Bigl \{\coth(\omega/2 k_B T) [\cos(\omega t)-1] 
 - i \sin(\omega t) \Bigr \} 
\end{equation}
for a given spectral density $I(\omega)$ of electromagnetic modes.
As a probability, $P(E)$ is normalized,
\begin{equation} \label{normp}
\int_{-\infty}^\infty P(E) = 1
\end{equation}
and fulfills the detailed balance relation, $P(-E)=\exp(-E/k_B T) P(E)$.

For clarity, we now focus on the most interesting zero-temperature case.
Provided $I(\omega)$ remains finite for low frequencies,
Eqs.~(\ref{tdossct}) and (\ref{jf}) straightforwardly lead to
a power law for the TDOS with exponent $\alpha = I(\omega\to 0)$.
We then should establish the harmonic nature
of the electromagnetic modes and compute the low-frequency
spectrum $I(\omega)$.  If $I(\omega\to 0)$ is finite,
a power law would directly follow.
Notice that a perturbative treatment of 
interactions is apparently not sufficient, as
a power-law TDOS is inconsistent
with conventional perturbative (1D or 2D) AAL
predictions \cite{altshuler}. 
In some studies \cite{sc3,finn}, $I(\omega)$ is phenomenologically 
parameterized in terms of the total impedance $Z(\omega)$, i.e.~the
MWNT is modelled as a transmission line.  Under such an
approach, one obtains $\alpha= Z(0)/(h/2e^2)$.
This simple transmission line model can directly explain
the doubling of the end exponent, because in the bulk
case one has effectively two resistances in parallel as
compared to the end case.   However, since this purely
phenomenological approach can hardly represent a satisfactory 
theory,  we pursue a {\sl microscopic}
approach.  It should also be
stressed at this point that the 1D pseudo-gap TDOS
found for small $E$ \cite{prl,gl1} is apparently
outside the reach of transmission line modelling.
Furthermore, the doubling of the end exponent
can be verified from microscopic theory as well.
The derivation is discussed at length in Ref.~\cite{prl}, and here
we focus on the bulk TDOS alone.

Recent field-theoretical  developments 
allow to incorporate the Coulomb interactions
in a nonperturbative way \cite{finkel,kopietz,ka}.
Adapting the Keldysh nonlinear $\sigma$ model approach for
interacting disordered systems worked out in Ref.~\cite{ka}, 
the TDOS can be computed in analytical form
for arbitrary interaction strength. 
This calculation is certainly on sound footing for long-ranged interactions
(which is the case for MWNTs) in 2D.  For
truly 1D systems, however, the asymptotic low-energy
behavior of the TDOS resulting from this approach, see Ref.\cite{gl1}, 
has been questioned recently \cite{kopietz}.  
In any event, the final result
yielded by this theory indeed reproduces phenomenological Coulomb blockade
theory, since the electromagnetic modes are found to be Gaussian 
with spectral density  \cite{prl,ka}
\begin{equation} \label{iw} 
I(\omega) = \frac{\omega}{\pi} {\rm Im}
\sum_{\vec{q}} \frac{1}{(Dq^2 - i\omega)^2} 
 \left( U_0^{-1}(\vec{q}) +
\frac{\nu_0 Dq^2}{Dq^2-i\omega}\right)^{-1} \;.
\end{equation}
Here the diffusion coefficient for 
charge diffusion on the tube surface is $D=v_F^2 \tau/2$.
In Eq.~(\ref{iw}), the $\vec{q}$ summation includes 
an integral over the momentum
parallel to the tube direction (a very long MWNT is assumed), 
and a summation over the discrete transverse momenta $q_\perp = n/R$
for integer $n$. For consistency, 
the $n$ summation is restricted to $|n|\leq N$,
albeit the detailed value for the cut-off is not essential.
 The Fourier-transformed Coulomb
interaction potential  $U_0(\vec{q})$ includes the effect
of external screening by nearby gates or the substrate,
but {\sl not}\ of internal screening which is fully accounted for by
Eq.~(\ref{iw}).  In what follows, to keep the discussion simple, we
consider an effectively short-range
interaction potential characterized by a constant $U_0$; for the 
case of a $1/r$ potential, see, e.g.~Ref.~\cite{ka}.
Since the dominant contributions to $I(\omega)$ come from small
$q$, it is justified to integrate
 over the longitudinal momentum in Eq.~(\ref{iw})
directly (there is no UV divergence), leading to
\begin{eqnarray} \label{iw1}
I(\omega) &=& \frac{U_0 }{2\pi (D^\ast-D)}  \, {\rm Re} \,
\sum_{n=-N}^N  \Bigl [
(-i\omega/D^\ast + n^2/R^2)^{-1/2}  \\ \nonumber && \quad -
(-i\omega/D + n^2/R^2 )^{-1/2} \Bigr ]
\end{eqnarray}
with the field diffusion constant $D^*=D (1+\nu_0 U_0)$.  
Although we do not present the derivation of Eq.~(\ref{iw}) here,
we feel it is important to summarize the main approximations 
entering this result:
\begin{enumerate}
\item
The regime $\ell\lapx 2\pi R$ is considered, but 
 Eq.~(\ref{iw1}) should also yield
useful results for somewhat larger $\ell$, since then
the $n=0$ mode dominates the Coulomb blockade completely.  The
$n=0$ mode is unaffected by assumptions concerning transversal motion.
\item
As a consequence of the assumed diffusive
behavior, the spectral density (\ref{iw1})
should only be used for $\omega \tau \ll 1$.
The $I(\omega\to 0) \sim \omega^{-1/2}$ 
behavior in Eq.~(\ref{iw}) due to the $n=0$ mode
implies that the dominant contribution 
to the Coulomb blockade indeed results from these
low-energy collective modes.  In Eq.~(\ref{jf}) we therefore truncate the
integration at the upper limit $1/\tau$. Since the higher energy 
modes are equivalent to a Luttinger liquid, which in turn 
only leads to comparatively weak Coulomb blockade effects
in this regime, their omission
is not expected to create serious problems.  
\item
The Coulomb interaction potential should be 
sufficiently long-ranged and smooth to
allow for semi-classical (WKB-type) treatments.  The main
effect of the interaction is then to change the phase of 
electron wavefunctions but not the amplitude.
Since Coulomb blockade is a low-energy 
collective phenomenon and the interaction
potential is rather long-ranged in MWNTs, 
this approximation should be justified.
\item
Only the intrinsic electrodynamic modes
 of the MWNT are considered to be responsible 
for the Coulomb blockade, but not 
the attached external circuit.  For sufficiently long MWNTs 
such as the ones in Ref.~\cite{sc3},  the intrinsic
resistance is large and
 environmental Coulomb blockade can safely be neglected.
\end{enumerate}

Above the Thouless energy for diffusion around the 
circumference, 
\begin{equation} \label{thouless}
E_T \approx \frac{D}{(2\pi R)^2} \;,
\end{equation}
where charge diffusion is essentially two-dimensional,
we then expect basically an exponentiated 2D AAL law.
In contrast, for sufficiently low energy scales, 
$E < E_T$, 1D behavior takes over, where 
the $n=0$ mode in Eq.~(\ref{iw1}) becomes more and more
important as the energy scale is decreased.
In these two limits one can obtain the TDOS analytically.  
In the {\sl 2D diffusive energy regime} $E > E_T$, a power
law emerges by converting the $n$-summation into an integral.
The bulk exponent  follows directly as $\alpha=I(\omega\to 0)$, and reads
\cite{prl}
\begin{equation} \label{alph}
\alpha = \frac{R}{2\pi \nu_0 D} \ln(D^*/D) \;.
\end{equation}
In effect, the AAL logarithmic correction therefore {\sl exponentiates}
into a {\sl power law} \cite{prl}.  We wish to stress that the
derivation of Eq.~(\ref{alph}) works only in the true 2D limit, 
characterized by a large number of bands $M$ or by $\ell \ll R$. 
On the other hand, for sufficiently low energy scales, only the $n=0$
mode responsible for 1D perturbative AAL corrections is important.
Keeping only the $n=0$ term in Eq.~(\ref{iw1}) 
results in a divergent (``sub-Ohmic'') behavior of
the low-frequency spectral density, $I(\omega)\sim \omega^{-1/2}$,
and hence to the appearance of a pseudo-gap as $E\to 0$ \cite{weiss},
\begin{equation} \label{1dlimit}
\nu(E) \sim \exp(-E_0/E) \;,
\end{equation}
where we neglect a prefactor exhibiting
 power-law energy dependence.  Alternatively, this leads to
logarithmic corrections of the scale $E_0$ in Eq.~(\ref{sce0}) below.
The result (\ref{1dlimit}) agrees with the findings of Refs.~\cite{gl1,roll}.
Using a stationary-phase evaluation
of $P(E)$ in Eq.~(\ref{pe}),
the energy scale $E_0$ in Eq.~(\ref{1dlimit}) follows as \cite{weiss}
\begin{equation}\label{sce0}
E_0  = \frac{U^2_0}{8\pi D \left(1+\sqrt{D^\ast/D}\right)^2 } \;,
\end{equation}
For strong interactions, $D^\ast\gg D$, this can be simplified
to $E_0= U_0/8\pi \nu_0 D$.
Remarkably, Eq.~(\ref{1dlimit}) representing the exponentiated 1D AAL law
does {\sl not}\ reproduce the perturbative 1D AAL $1/\sqrt{E}$ behavior under
a naive direct expansion of the exponential \cite{gl1,roll}. 
 
\section{Numerical solution of Coulomb blockade equations}
\label{sec4}

To analyze the full energy dependence of the TDOS,
numerical methods are mandatory.  Using the spectral density (\ref{iw1}),
the phase correlation function $J(t)$ in Eq.~(\ref{jf}) can
be evaluated, which then allows for the computation of $P(E)$
according to Eq.~(\ref{pe}).  Finally, Eq.~(\ref{tdossct}) yields
the TDOS.  For simplicity, we again focus on the $T=0$ 
limit, although the finite-temperature case is also directly
accessible.  A convenient check that the  numerical procedure 
has converged is provided by
the normalization condition for $P(E)$, see Eq.~(\ref{normp}),
which is accurately fulfilled for the results reported below. 
For convenience,
the energy scale is set by $v_F/R$ throughout 
this section. We consider a situation
with $k_F R=5.5$ so that $N=5$ and the number of bands is $M=11$, and 
 use the value
$U_0/2\pi v_F = 1$ to parameterize the interaction strength.
Unfortunately, it appears to be rather difficult to compute a realistic value
for $U_0$ (except possibly by ab-initio methods).
The above choice corresponds to rather strong interactions, 
but for comparison
we have also carried out calculations 
for $U_0/v_F=1$ (not shown, but see below).

\begin{figure}
\epsfysize=10cm
\epsffile{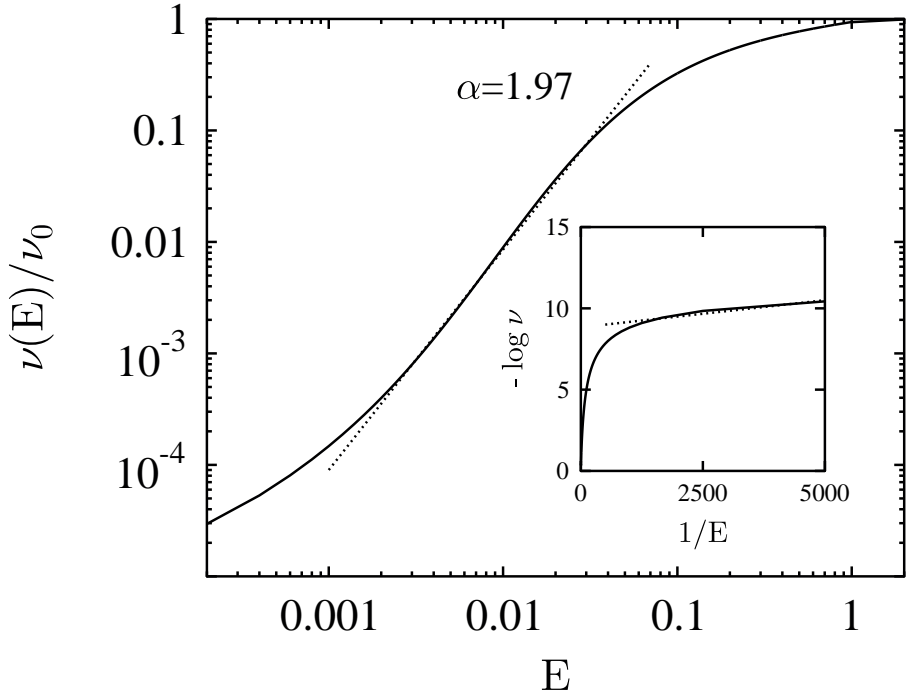}
\caption{ \label{fig1}
Numerical result for the energy-dependence of the TDOS at $\ell=R$
on a double-logarithmic scale (for other parameters, see text).
The dotted line represents a power-law fit with exponent $\alpha=1.97$.
Inset: Arrhenius plot of the same data.  The dotted line
has slope $E_0=0.015$.}
\end{figure}

\begin{figure}
\epsfysize=10cm
\epsffile{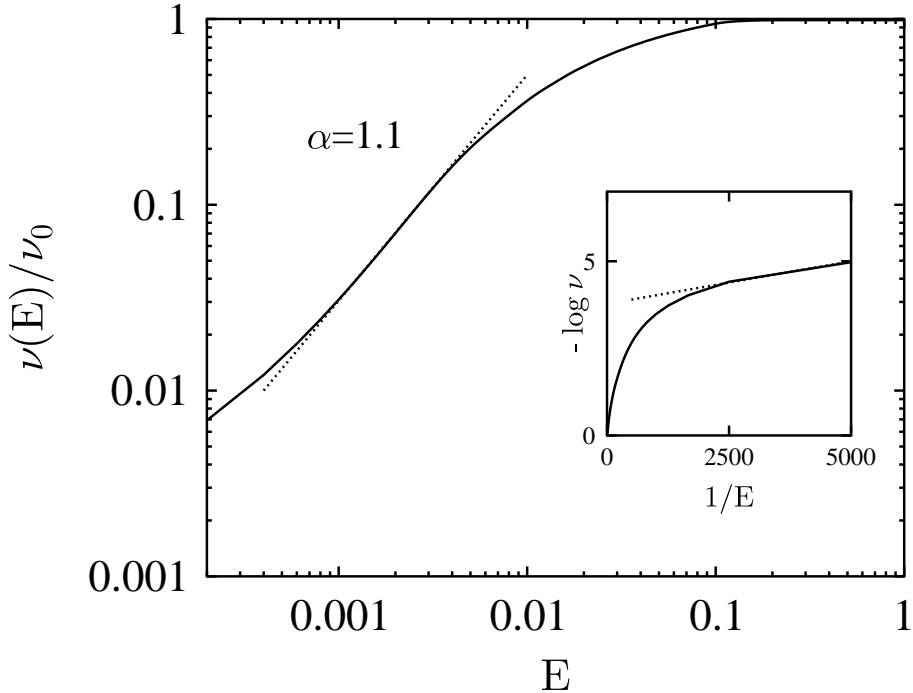}
\caption{ \label{fig2}
Same as Fig.~\ref{fig1} but for $\ell=10 R$.
The dotted line is a power-law fit with exponent $\alpha=1.1$.
Inset: Arrhenius plot of the same data.  The dotted line
has slope $E_0=0.00025$.}
\end{figure}

Let us start with a rather dirty MWNT,  $\ell = R$.
In the mentioned units, 
the Thouless scale is $E_T= 0.013$, and the
plot of the TDOS in Fig.~\ref{fig1} indeed shows an apparent {\sl power law} 
even for energies well below $E_T$, extending up to $E\approx 0.1$ 
over approximately one decade.
The inset shows that for $E\to 0$, the predicted pseudo-gap behavior 
emerges.  For this parameter set,
the power-law exponent $\alpha=0.23$ predicted 
from the 2D limit, Eq.~(\ref{alph}), is much smaller than the
numerically observed exponent $\alpha=1.97$. 
The estimate (\ref{alph}) is therefore too crude and really restricted
to the true 2D regime, since it
ignores the special role played by the $n=0$ contribution in
the spectral density. 
Nevertheless, we find a clear power-law behavior at intermediate
energy scales.  Importantly, the 
regime of validity for the power law is not set by the Thouless scale,
but by a smaller energy scale.  
Similarly, the value $E_0=0.015$ extracted from the slope of the 
Arrhenius plot in the inset
of Fig.~\ref{fig1} is significantly smaller than the 
 value $E_0=0.07$ predicted by Eq.~(\ref{sce0}).  
This deviation is probably linked to strong logarithmic renormalizations
of the scale (\ref{sce0}) by the power-law prefactor
not written out in Eq.~(\ref{1dlimit}), and is always observed in
our calculations.
Since this renormalization of $E_0$ makes this scale quite
small, the pseudo-gap TDOS  is
in practice difficult to distinguish from the power law
except at very low energies.   This may offer a (somewhat trivial)
explanation for the experimental difficulties encountered in finding
pseudo-gap behavior.  Finally, for high energies
 close to $v_F/R$, the non-interacting DOS is approached.

Next we discuss the {\sl quasi-ballistic limit}, taking $\ell=10 R$.
The features shown in Fig.~\ref{fig2}
are qualitatively similar as in Fig.~\ref{fig1},
namely a pseudo-gap at very low energies turns into a power law at intermediate
energies.  The power law exponents become systematically smaller by increasing
$\ell$, in this case $\alpha=1.1$.
The power law crosses over into the pseudo-gap as $E\to 0$, with $E_0=0.00025$
again much smaller than the  expected value $E_0=0.007$.
As this power law feature is definitely not linked to the 
Thouless scale, it is not related to the exponent (\ref{alph})
for tunneling into a 2D interacting disordered metal.
We have checked for this value of $\ell$ that the 
power law persists for smaller $U_0$. In fact, the 
exponent $\alpha$ then systematically decreases, and
for $U_0/v_F=1$ is $\alpha\approx 0.3$, which would be in
good agreement with experiment.
Since the interaction strength in the experiments conducted in Ref.~\cite{sc3}
did probably not vary much from tube to tube, the robustness of
the observed exponents with respect to changes in $\ell/R$
is encouraging.  We mention in passing that 
for suspended  MWNTs or smaller doping levels, one may 
reach a regime of stronger interactions, where again
power-law behavior at 
intermediate energies is predicted, but with larger exponents.

\section{Conclusions}
\label{sec5}

MWNTs represent a unique laboratory for exploring mesoscopic
physics in the presence of electron-electron interactions.
Here we have addressed one aspect, namely the zero-bias
anomaly of the tunneling density of states due to 
Coulomb interactions among the electrons.
Assuming a sufficiently dirty MWNT with mean free path
less than the circumference,
the spectral density $I(\omega)$ of Coulomb blockade theory
has been computed, and the numerical
solution of the resulting equations for the TDOS was presented.  The results 
show power-law behavior at intermediate energy scales,
extending down to quite low energies over typically one to
two decades. Remarkably, the power law is seen at
energies less than the Thouless
scale for diffusion around the circumference, i.e.~it does
not appear to reflect tunneling into a 2D diffusive electron
liquid.  At very low energies,
the power-law behavior $\nu(E) \sim E^\alpha$
crosses over into a pseudo-gap of the
form $\nu(E) \sim \exp(-E_0/E)$, although it
should be stressed that in practice both are 
sometimes hard to distinguish
because of the rather small scale $E_0$.
According to our numerical study,
the power-law exponent $\alpha$ would be consistent with
typical exponents around $\alpha\approx 0.3$
for interaction strength $U_0/v_F\approx 1$,
but is significantly larger for stronger interactions. 
We note that the power law is very robust, and has always been
observed in our calculations, regardless of the chosen values
for $\ell/R$ or $U_0$.  

Support by the DFG under the Gerhard-Hess program and by the
EPSRC under Grant No.~GR/N19359 is gratefully acknowledged.

\end{document}